\documentclass[secnumarabic,showpacs,amsmath,amssymb,prd]{revtex4}

\begin{document}

\title{A cosmology with variable c}

\author{Hossein Shojaie}
 \email{h-shojaie@sbu.ac.ir}

\author{Mehrdad Farhoudi}
 \email{m-farhoudi@sbu.ac.ir}

\affiliation{Department of Physics, Shahid Beheshti University,\\
Evin, Tehran 1983963113, Iran}

\begin{abstract}
A new varying-$c$ cosmological model constructed using two additional
assumptions, which was introduced in our previous work, is briefly
reviewed and the dynamic equation of the model is derived distinctly
from a semi-Newtonian approach. The results of this model, using a
$\Lambda$ term and an extra energy-momentum tensor, are considered
separately. It is shown that the Universe began from a hot Big Bang
and expands forever with a constant deceleration parameter
regardless of its curvature. Finally, the age, the radius, and the
energy content of the Universe are estimated and some discussion
about the type of the geometry of the Universe is provided.
\end{abstract}

\pacs{98.80.Bp, 98.80.Jk}

\maketitle

\section{Introduction}

During the last decade, varying speed-of-light (VSL) theories
have arisen as alternatives to the inflationary scenario of the
Universe~\cite{Moffat-1993,Albrecht-Magueijo-1999,
Barrow-Magueijo-1998,Clayton-Moffat-1998,Barrow-1999,
Barrow-Magueijo-1999,Clayton-Moffat-2000,Magueijo-2000,Basset-2000}.
These theories are based on various dynamics but in all of them,
the speed of light needs be much more larger than at present to
solve the cosmological problems. For instance, in a model provided
by Albrecht and Magueijo~\cite{Albrecht-Magueijo-1999}, the speed
of light suddenly falls off to its current value during a phase
transition. Some authors introduced bimetric gravity
theories~\cite{Clayton-Moffat-1998,Clayton-Moffat-2000}. Barrow
and Magueijo constructed a model in which the speed of light
decreases by a smooth power-law function of the cosmological
time~\cite{Barrow-Magueijo-1998}.

There are some criticisms about the varying-$c$ models. For
instance, the speed of light is not a dimensionless quantity,
hence, going to a new frame, one may cancel its probable
variations. Or, if $c$ and, consequently, the coupling constant of
the Einstein equation would vary, then observers in different
frames would see the evolution of the Universe governed by
different rules. However, these arguments are also applicable to
any varying-constant model, in which some physical constants are
made to vary, for example, any theory with varying gravitational
constant $G$, such as the Brans-Dicke theory of gravitation. Moreover,
if a dimensionless parameter seems to vary, for example, the fine-structure
constant $\alpha$, then its included quantities, which
are dimensional parameters, are to vary as well. In this case, if
one tries to fix the value of these latter quantities, one simply
lets the units be rescaled.

Some authors worry about the Lorentz invariance breaking in varying speed-of-light models.
In contrast, the authors of ref.~\cite{Jafari-Shariati-2003} claim that they
have provided the Fock-Lorentz~\cite{Fock-1964,Manida-1999} and
Magueijo-Smolin~\cite{Magueijo-Smolin-2002,Magueijo-Smolin-2003}
transformations with a varying-$c$, as redescriptions of
special relativity. Also, according to ref.~\cite{Rindler-1979},
assuming only the first principle of special relativity, namely,
the relativity principle, together with Euclidicity and isotropy
leads to either Galilean or Lorentz transformations. In the latter
transformation, there is an upper local limit for the speed of
particles. In addition, the semi-strong principle of equivalence permits the
possibility of different numerical contents, namely, the values of
the fundamental constants, at different parts of space-time in
the Universe. The global laws, whose local approximations are
considered in the various local space-time regions, may probably
involve the derivatives of these constants~\cite{Rindler-1979}.

Other objections are due to the fairly well-predicted
nucleosynthesis and cosmic microwave background (CMB). Till now, no
one has claimed that the varying-constant models cannot predict
these phenomena.

At last, some people prefer the frame-independent arguments. It is
a plausible opinion in every field of physics except
cosmology, for when one considers the Universe as a whole, one cannot
ignore the preferred frame of cosmology, or introduce a more
useful frame than it; that is, it is of worth for a law to be
independent of the observers so as to be identical everywhere,
but the cosmological preferred frame is the most global frame
that is obtainable everywhere by its homogeneity and isotropy.
Moreover, in a Machian view, this frame is an undetachable
property of the Universe.

In our previous work~\cite{Shojaie-Farhoudi-2004-1}, we showed that
one can construct a simple varying-$c$ model of the Universe using two
key assumptions, which we will review in Sect.~\ref{Brief0}. In
addition, this model is also based on the cosmological principle, the Weyl
postulate, and the extended Einstein equation, where, by the latter, we
mean that the coefficient of the Einstein equation is a function of
cosmological time in the preferred cosmological frame. This
model provides solutions to some of the Standard Big Bang (SBB)
problems. It should be mentioned that the consequent results were
obtained from two parts: the first part was directly derived from
our assumptions that were discussed as the basic concepts.
Section~\ref{Brief0} gives a brief review of this part. The second
part provides the dynamics of the Universe, i.e., the Friedmann
equations.

In this work, we are going to consider a different approach to
deriving those Friedmann equations, namely, a semi-Newtonian way.
Also, a comparison of the results of this model with the Einstein
equation when it is equipped with the cosmological ``constant'', $\Lambda$,
is provided in Sect.~\ref{Lambda0}.
Section~\ref{Tensor0} includes a similar way of deriving the
Friedmann equations, where $\Lambda$ is replaced by an extra
energy-momentum tensor related to the varying-$c$. Finally, in
Sect.~\ref{Review0}, the results of this model are discussed,
some ideas about the age and the radius of the Universe are provided
and the energy content of it is estimated. In addition, by means
of the derived values, the type of the geometry of the Universe is
discussed.

\section{Brief review}\label{Brief0}

The very plausible assumptions of homogeneity and isotropy of the
Universe, which are in agreement with observations on the cosmic
scale, lead one to use the Robertson-Walker metric as the
preferred frame for cosmology. In general, one can write this
metric as
\begin{equation}
ds^2=c(t)^2dt^2-a(t)^2\left(\frac{dr^2}{1-kr^2}+r^2d\Omega^2\right)\label{Brief15}
\end{equation}
which is conformal with its usual form, i.e.,
\begin{equation}
ds^2=\left(\frac{c(t)}{c_0}\right)^2\left[c_0^2dt^2-a'(t)^2
\left(\frac{dr^2}{1-kr^2}+r^2d\Omega^2\right)\right]\label{Brief16}
\end{equation}
From now on, we hold all arguments in the preferred frame of cosmology and
write them according to the metric~(\ref{Brief15}). This is the preferred metric
for the whole Universe and not its constituents.

To construct the model, the first assumption is that the total
energy of the Universe, as observed in the preferred frame of the
cosmology, is a constant, i.e.,
\begin{equation}
Mc^2=const.\label{Brief1}
\end{equation}
where $M$ is the total mass of the Universe. This is what has
really been accepted in the SBB. However, if one ignores the
constancy of the speed of light, then the situation
differs completely from the SBB.

The second assumption asserts that the total energy of a particle,
including its inertial energy and its gravitational potential
energy, measured in the preferred frame of the cosmology, is zero.
Mathematically, that is
\begin{equation}
-\frac{GMm}{R}+mc^2=0\label{Brief2}
\end{equation}
where $R$ is the radius of the Universe and $G$, as we
assume, is a constant. In other words, the inertial energy of a
particle is due to the gravitational potential energy of the mass
content of the Universe upon it. Simplifying the above relation,
it gives
\begin{equation}
\frac{GM}{Rc^2}=1\label{Brief3}
\end{equation}

This dimensionless relation is a legitimate one that contains the
most essential cosmological quantities such as $M$, $R$ and $c$.
Inspired from this, one may feel better if one substitutes the
second assumption as if one defines $G$ as a proportionality
coefficient with which the relation~(\ref{Brief3}) holds. That is,
the Universe is such that the fraction of $\frac{Rc^2}{M}$ is a
constant.

Combining the relations~(\ref{Brief1}) and~(\ref{Brief3}),
to omit $M$, and then differentiating the result, one gets
\begin{equation}
\frac{dc}{c}=-\frac{1}{4}\frac{dR}{R}\label{Brief4}
\end{equation}
Now, using the ordinary dimensionless scale factor, $a$, as
\begin{equation}
\frac{a}{a_0}=\frac{R}{R_0}\label{Brief5}
\end{equation}
one can gain from the relation~(\ref{Brief4}) that
\begin{equation}
c=c_0\left(\frac{a}{a_0}\right)^{-\frac{1}{4}}\label{Brief6}
\end{equation}
where $c_0$ and $a_0$ are the current values of these quantities.
Following the same procedure in deriving the above equation, one
can easily get
\begin{equation}
M=M_0\left(\frac{a}{a_0}\right)^\frac{1}{2}\label{Brief7}
\end{equation}
The relations~(\ref{Brief5}) and~(\ref{Brief7}) imply that:
\begin{equation}
\rho\propto \frac{M}{R^3}\propto \frac{1}{a^\frac{5}{2}}\quad
\textrm{
or}\quad\rho=\rho_0\left(\frac{a}{a_0}\right)^{-\frac{5}{2}}\label{Brief8}
\end{equation}
where $\rho$ is the total matter density of the Universe,
regardless of its content. Combining the relations~(\ref{Brief6})
and~(\ref{Brief8}), one gets
\begin{equation}
\varepsilon=\rho c^2=\rho_0{c_0}^2\left(\frac{a}{a_0}\right)^{-3}\label{Brief10}
\end{equation}
which is a more familiar relation.

We also found in ref.~\cite{Shojaie-Farhoudi-2004-1} that the
wavelength and the frequency of the light vary as
\begin{equation}
\lambda=\lambda_0\frac{a}{a_0}\quad\textrm{ and}\quad
f=f_0\left(\frac{a}{a_0}\right)^{-\frac{5}{4}}\label{Brief11}
\end{equation}
In addition, the relation between the emitted frequency, $f_e$,
the received frequency, $f_r$, and the redshift, $z$, is
\begin{equation}
\frac{f_e}{f_r}=\left(\frac{a_0}{a_z}\right)^\frac{5}{4}=\left(1+z\right)^\frac{5}{4}\label{Brief13}
\end{equation}
where $a_z$ is the value of the scale factor at the time of
emission. Also, we found that the temperature of the Universe
varies as
\begin{equation}
T=T_0\left(\frac{a}{a_0}\right)^{-\frac{5}{4}}\label{Brief14}
\end{equation}

\section{Semi-Newtonian approach}\label{Semi0}

A semi-Newtonian approach to the dynamics of the Universe, often used
in standard text books~\cite{Harrison-2000,Liddle-2003},
is mainly based on Newtonian gravitation (except the existence of
any absolute space or time) instead of the general relativity.

Using the assumptions of homogeneity and isotropy, one can write the energy equation for
a test particle $m$ at a distance $r$ away from any center, which
can be selected arbitrarily, as
\begin{equation}
U=\frac{1}{2}m\dot
r^2-\frac{Gm\left(\frac{4\pi}{3}r^3\rho\right)}{r}\label{Semi1}
\end{equation}
Rearranging the equation, using the relation $\vec r=a(t)\vec x$,
where $\vec r$ and $\vec x$ are related as physical and
co-moving coordinates, respectively, one gets
\begin{equation}
\frac{2U}{mx^2}=\dot a^2-\frac{8\pi}{3}Ga^2\rho\label{Semi2}
\end{equation}
The right-hand side of the above relation does not depend on
spatial coordinates and consequently neither does the left-hand side, so
in fact, one learns that homogeneity requires that $U$ does indeed
vary with $x^2$. In addition, inspired by special
relativity, one has $\frac{U}{m}\propto c^2$. These lead to
\begin{equation}
kc^2\equiv-\frac{2U}{mx^2}=-\dot a^2+\frac{8\pi}{3}Ga^2\rho\label{Semi6}
\end{equation}
where $k$ is supposed to be a constant and for further
convenience, it should be rescaled to unity. Also, it is worth
noting that the second term on the right-hand side of
relation~(\ref{Semi2}), regarding the relations~(\ref{Brief6})
and~(\ref{Brief8}), is proportional to the square of the speed of
light as well. Therefore, one can write
\begin{equation}
\left(\frac{\dot a}{a}\right)^2+\frac{kc^2}{a^2}=\frac{8\pi
G}{3}\rho\label{Semi3}
\end{equation}
which is exactly the standard form of the first Friedmann
equation. By differentiating this relation with respect to $t$,
one gets
\begin{equation}
2\left[\frac{\ddot a}{a}-\left(\frac{\dot a
}{a}\right)^2\right]\frac{\dot
a}{a}+2\frac{kc^2}{a^2}\left(\frac{\dot c}{c}-\frac{\dot
a}{a}\right)=\frac{8\pi G}{3}\dot\rho\label{Semi4}
\end{equation}
Using the relations~(\ref{Brief6}) and~(\ref{Brief8}), one gets
\begin{equation}
\frac{\ddot a}{a}=-\frac{1}{4}\left(\frac{\dot a}{a}\right)^2\label{Semi5}
\end{equation}
which is the analog of the second Friedmann equation in this
model. The other equations can be derived from this point on,
namely, an ever-expanding Universe, i.e.,
\begin{equation}
a=a_0\left(\frac{t}{t_0}\right)^{\frac{4}{5}}\label{Review1}
\end{equation}
regardless of the type of the geometry of the Universe, with
\begin{equation}
t_0=\frac{4}{5}\frac{1}{H_0}\label{Review23}
\end{equation}
where $H_0$ is the value of the Hubble constant, i.e.,
$H=\frac{\dot a}{a}$ at $t_0$. Consequently, one has
\begin{equation}
c(t)=c_0\left(\frac{t}{t_0}\right)^{-\frac{1}{5}}\ ,\quad
M(t)=M_0\left(\frac{t}{t_0}\right)^\frac{2}{5}\
,\quad\rho(t)=\rho_0\left(\frac{t}{t_0}\right)^{-2}\ ,\quad
T=T_0\left(\frac{t}{t_0}\right)^{-1}\label{Review3}
\end{equation}

However, one should notice that the assumption~(\ref{Semi6}) in
the Newtonian view is actually a priori demand for obtaining
the relation~(\ref{Review1}). Also, the time dependency of
$\frac{U}{m}$ cannot be seen in the classical way. Hence, because of
these points, it is a semi-Newtonian approach.

\section{The field equation with $\Lambda$}\label{Lambda0}

Although it is not necessary to introduce a $\Lambda$ term in the
Einstein equation when treating this model, for similarity to
the known models, in this section, we treat the model as if it has
two sources, the ordinary energy-momentum tensor plus a $\Lambda$
term, namely,
\begin{equation}
G^{ab}=\frac{8\pi G}{c^4}T_m^{ab}+\Lambda g^{ab}\label{Lambda1}
\end{equation}
The index $m$ in $T_m^{ab}$ has been introduced in order to
distinguish it from the energy-momentum tensor, $T^{ab}$,
introduced in ref.~\cite{Shojaie-Farhoudi-2004-1}, where
$\nabla_aT^{ab}\neq0$. Now, by $T_m^{ab}$, we mean the
energy-momentum tensor of a part of the matter for which the
conservation still holds, i.e. we assume
\begin{equation}
\nabla_aT_m^{ab}=0\label{Lambda3}
\end{equation}
In the above, we regard the $\Lambda$ term as a source of
matter than geometry. Also, as the properties of the
Robertson-Walker metric are based on the symmetries of
space-time, we use its adapted frame as the preferred coordinate,
hence, the components of this metric are
\begin{equation}
g_{00}=1\quad ,\quad g_{11}=-\frac{a^2(t)}{1-kr^2}\quad ,\quad
g_{22}=-a^2(t)r^2\quad ,\quad g_{33}=-a^2(t)r^2\sin ^2\theta\label{Lambda15}
\end{equation}
As is obvious, the metric components are as before, and
this allows us to use the well-known Friedmann equations with $x^0$,
and then one replaces $dx^0$ by $cdt$, see
ref.~\cite{Shojaie-Farhoudi-2004-1} for details.

The inner property of the geometry, i.e. the reduced Bianchi
identity, $\nabla_aG^{ab}=0$, implies that
\begin{equation}
\nabla_a\left(\frac{8\pi G}{c^4}T_m^{ab}+\Lambda g^{ab}\right)=0\label{Lambda2}
\end{equation}
If further, one assumes $T_m^{ab}$ to be a perfect fluid with the
equation of state as
\begin{equation}
p_m=(\gamma_m-1)\rho_m c^2\label{Lambda16}
\end{equation}
then, one can simply solve the relation~(\ref{Lambda3}) to get
\begin{equation}
\rho_m \propto \frac{1}{a^{3\gamma_m-1/2}}\quad \textrm{ and
hence}\quad \rho_m c^2 \propto \frac{1}{a^{3\gamma_m}}\label{Lambda4}
\end{equation}
or
\begin{equation}
\rho_m={\rho_m}_0\left(\frac{a}{a_0}\right)^{\frac{1}{2}-3\gamma_m}\label{Lambda5}
\end{equation}
where ${\rho_m}_0$ is the matter density related to $T_m^{ab}$ at
$a_0$. Using the relation~(\ref{Lambda3}) in the
relation~(\ref{Lambda2}) yields
\begin{equation}
\partial_a\left(\frac{8\pi G}{c^4}\right)T_m^{ab}+\left(\partial_a\Lambda\right)g^{ab}=0\label{Lambda6}
\end{equation}
Hence, as $c$ is not a constant, $\Lambda$ cannot be a constant
anymore, and it must be a function of space-time in general.
However, in this case, from the Lagrangian point of
view~\cite{Lovelock-Rund-1975}, for any Lagrangian density that
is a function of the metric and its first and second derivatives,
where its functionality on the second derivatives of the metric is
linear, a varying $\Lambda$ should be related to the curvature
scalar. However, we are not going to proceed any further with this view in this
work.

By the homogeneity and in the preferred frame of the cosmology,
which is being used here, $\Lambda$ is just a function of
cosmological time only. Hence, the zero-component of the
relation~(\ref{Lambda6}) gives
\begin{equation}
8\pi G\partial_0\left(\frac{1}{c^4}\right)T_m^{00}+
\partial_0\Lambda g^{00}=0\label{Lambda7}
\end{equation}
Using relation~(\ref{Brief6}) yields
\begin{equation}
\Lambda=\Lambda_0\left(\frac{a}{a_0}\right)^{1-3\gamma_m}\label{Lambda8}
\end{equation}
where
\begin{equation}
\Lambda_0=\frac{8\pi G{\rho_m}_0}{(3\gamma_m-1)c_0^2}\label{Lambda9}
\end{equation}
This reminds us the quintessence instead of the cosmological
``constant''.

To obtain $\gamma_m$, one should use the zero-component of~(\ref{Lambda1})
to get the first Friedmann equation, i.e.,
\begin{equation}
\left(\frac{\dot a}{a}\right)^2+\frac{kc^2}{a^2}=\frac{8\pi
G}{3}\rho_m+\frac{\Lambda c^2}{3}\label{Lambda10}
\end{equation}
From the relations~(\ref{Brief6}),~(\ref{Lambda5})
and~(\ref{Lambda8}), it is clear that the two terms on the
right-hand side of the above equation have the same order of the
scale factor. So, if one compares~(\ref{Semi3}) with
the above relation, and then uses relation~(\ref{Brief8}), one gets
\begin{equation}
\frac{8\pi G}{3}\rho_m+\frac{\Lambda c^2}{3}\propto
a^{-\frac{5}{2}}\label{Lambda11}
\end{equation}
which directly implies
\begin{equation}
\gamma_m=1\label{Lambda12}
\end{equation}
The above result corresponds to dust, i.e., $p_m=0$. This
simplifies the relations~(\ref{Lambda8}) and~(\ref{Lambda9}) as
\begin{equation}
\Lambda=\Lambda_0\left(\frac{a}{a_0}\right)^{-2}\label{Lambda13}
\end{equation}
where
\begin{equation}
\Lambda_0=\frac{4\pi G}{c_0^2}{\rho_m}_0\label{Lambda14}
\end{equation}
and hence
\begin{equation}
\Lambda=\Lambda_0\left(\frac{t}{t_0}\right)^{-\frac{8}{5}}\label{Review7}
\end{equation}
It is worth noting that the splitting of the total mass of the
Universe into $T_m^{ab}$, the covariant divergence of which vanishes,
and a $\Lambda$ term that varies with the cosmological time, is
not unique. That is, $\rho_m$ is not fixed, and strictly speaking,
it depends on how one splits the mass and energy content of the
Universe between $\rho_m$ and $\Lambda$.

\section{The field equation with an additional energy-momentum
tensor}\label{Tensor0}

A plausible argument for considering the previous section is that
the term $\Lambda$ has been recently used in these types of
contexts. However, a better way to consider that section may be to
introduce a new energy-momentum tensor, $T_\phi^{ab}$. In other
words, one can split the total energy-momentum in the Einstein
equation, $G_{ab}=\frac{8\pi G}{c^4}T_{ab}$, as
\begin{equation}
T^{ab}=T_m^{ab}+T_\phi^{ab}\label{Tensor3}
\end{equation}
That is, to consider
\begin{equation}
G^{ab}=\frac{8\pi G}{c^4}\left(T_m^{ab}+T_\phi^{ab}\right)\label{Tensor1}
\end{equation}
where again, by $T_m^{ab}$ we mean
\begin{equation}
\nabla_aT_m^{ab}=0\label{Tensor2}
\end{equation}
Using the Robertson-Walker metric and $T_m^{ab}$ as a perfect
fluid, relation~(\ref{Lambda5}) is still valid. In addition,
for simplicity, we assume $T_\phi^{ab}$ to be a perfect fluid
too, with the usual equation of state
\begin{equation}
p_\phi=(\gamma_\phi-1)\rho_\phi c^2\label{Tensor4}
\end{equation}
Applying the covariant divergence to both sides of~(\ref{Tensor1}) leads to
\begin{equation}
\partial_a\left(\frac{1}{c^4}\right)\left(T_m^{ab}+
T_\phi^{ab}\right)+\frac{1}{c^4}\nabla_aT_\phi^{ab}=0\label{Tensor6}
\end{equation}
Simplification of the zero-component of the above relation, using
the Robertson-Walker metric, gives
\begin{equation}
\dot\rho_\phi+\left(3\gamma_\phi\frac{\dot a}{a}-2\frac{\dot
c}{c}\right)\rho_\phi-4\frac{\dot c}{c}\rho_m=0\label{Tensor7}
\end{equation}
Using the relation~(\ref{Brief6}) and assuming a plausible
assumption that $\rho_\phi\propto a^\ell$ results in
\begin{equation}
\rho_\phi=-\frac{1}{3\gamma_\phi+\frac{1}{2}+\ell}\rho_m\label{Tensor8}
\end{equation}
i.e., $\rho_m$ and $\rho_\phi$ are in the same order of $a$, and
consequently from~(\ref{Lambda5}), one gets
\begin{equation}
\ell=\frac{1}{2}-3\gamma_m\label{Tensor5}
\end{equation}
The relation~(\ref{Tensor1}) gives
\begin{equation}
\left(\frac{\dot a}{a}\right)^2+\frac{kc^2}{a^2}=\frac{8\pi
G}{3}\left(\rho_m+\rho_\phi\right)\label{Tensor9}
\end{equation}
and
\begin{equation}
\frac{\ddot a}{a}+\frac{1}{4}\left(\frac{\dot
a}{a}\right)^2=-\frac{4\pi
G}{3}\Big[(3\gamma_m-2)\rho_m+(3\gamma_\phi-2)\rho_\phi\Big]\label{Tensor10}
\end{equation}
The total density, $\rho$, calculated in
relation~(\ref{Brief8}) is the sum of the densities $\rho_m$ and
$\rho_\phi$. Hence, regarding relation~(\ref{Tensor8}), it
implies that
\begin{equation}
\rho_\phi\propto\rho_m\propto a^{-\frac{5}{2}}\label{Tensor11}
\end{equation}
Therefore, we get exactly
\begin{equation}
\rho_\phi=\frac{1}{2-3\gamma_\phi}\rho_m\label{Tensor14}
\end{equation}
where $\gamma_\phi\neq\frac{2}{3}$, and
\begin{equation}
\gamma_m=1\label{Tensor12}
\end{equation}
which, as in the previous section, shows that $T_m^{ab}$
corresponds to dust. Besides, using
\begin{equation}
p=p_\phi+p_m=p_\phi=(\gamma_\phi-1)\rho_\phi c^2\label{Tensor18}
\end{equation}
and
\begin{equation}
p=(\gamma-1)\rho c^2=(\gamma-1)(\rho_m+\rho_\phi)c^2\label{Tensor19}
\end{equation}
and relation~(\ref{Tensor14}) gives $\gamma=\frac{2}{3}$
as expected and as derived in
ref.~\cite{Shojaie-Farhoudi-2004-1}. Moreover, this is also
consistent with why $\gamma_\phi\neq\frac{2}{3}$.

Substituting relations~(\ref{Tensor14}) and~(\ref{Tensor12})
in relation~(\ref{Tensor10}) makes its right-hand side
vanish automatically and does not give a new relation. As
is discussed at the end of the Sect.~\ref{Lambda0}, $\rho_m$ is
not unique. In addition, relation~(\ref{Tensor14}) shows a
freedom to choose $\rho_\phi$ and $\gamma_\phi$, even when
$\rho_m$ is completely definite. That is, the value of
$\gamma_\phi$ cannot be fixed by these formulae and it should
perhaps be determined from its equation of state.

In a special case, when one sets
\begin{equation}
\rho_\phi=\frac{\Lambda c^2}{8\pi G}\equiv\rho_\Lambda\label{Tensor13}
\end{equation}
the relations~(\ref{Brief6}),~(\ref{Lambda13}),~(\ref{Lambda14}),~(\ref{Tensor11})
and~(\ref{Tensor14}) give
\begin{equation}
\gamma_\phi=0\label{Tensor15}
\end{equation}
which corresponds to a negative pressure
\begin{equation}
p_\phi=-\rho_\phi c^2=-\frac{1}{2}\rho_mc^2\label{Tensor16}
\end{equation}
or
\begin{equation}
p_\Lambda=-\rho_\Lambda c^2\ ,\label{Tensor17}
\end{equation}
of the previous section.

\section{Summary and discussion}\label{Review0}

We summarize the dynamics of the Universe and other
results of this model. The relations~(\ref{Review1})
and~(\ref{Review3}) imply that the Universe began from a hot Big
Bang and expands forever, even if $k=+1$. In addition, one has
from~(\ref{Semi5}) that
\begin{equation}
\frac{\ddot a}{a}=-\frac{1}{4}\left(\frac{\dot
a}{a}\right)^2=-\frac{1}{4}H^2\label{Review2}
\end{equation}
This relation asserts that the Universe always decelerates with a
constant deceleration parameter $q\equiv-\frac{\ddot aa}{\dot
a^2}=\frac{1}{4}$.

From measurements, the Hubble constant is given as
\begin{equation}
H_0=100h\frac{km}{s}\frac{1}{Mpc}\quad \textrm{ where}\quad
0.55\leq h\leq 0.8\label{Review26}
\end{equation}
Hence, by relation~(\ref{Review23}), the age of the Universe
should be
\begin{equation}
9.78 \times 10^9\ yr < t_0 < 14.22 \times 10^9\ yr\label{Review27}
\end{equation}
in this model. Generally, from relation~(\ref{Review1}), one gets
\begin{equation}
t=\frac{4}{5}\frac{1}{H}\label{Review25}
\end{equation}

One can also calculate the particle horizon as
\begin{equation}
\int^{r_0}_0\frac{dr}{\sqrt{1-kr^2}}=\int^{t_0}_0\frac{c(t)dt}{a(t)}=
\int^{t_0}_0\frac{c_0t_0}{a_0}\frac{dt}{t}=\frac{c_0t_0}{a_0}\ln
t\Big|^{t_0}_0\label{Review8}
\end{equation}
which is obviously infinite. Therefore, the observed radius of the
Universe is actually the current radius of the whole Universe.
This result means that all parts of the Universe are causally
connected and leads to an idea to estimate the current radius of
the Universe. From relations~(\ref{Brief5}),~(\ref{Review1})
and~(\ref{Review3}), one can get
\begin{equation}
dR=\frac{4}{5}R_0t_0^{-\frac{4}{5}}t^{-\frac{1}{5}}dt\label{Review9}
\end{equation}
and
\begin{equation}
cdt=c_0t_0^\frac{1}{5}t^{-\frac{1}{5}}dt\label{Review10}
\end{equation}
As can easily be seen, both the above relations have the same
time dependency, $t^{-\frac{1}{5}}$, in agreement with the above
assertion. To hold the causality, one can equate the
relations~(\ref{Review9}) and~(\ref{Review10}) and find
\begin{equation}
R_0=\frac{5}{4}c_0t_0\label{Review11}
\end{equation}
Substituting relation~(\ref{Review23}) in the above relation,
it yields
\begin{equation}
c_0=H_0R_0\label{Review24}
\end{equation}
or, using the
relations~(\ref{Brief5}),~(\ref{Review1}),~(\ref{Review23}),~(\ref{Review3})
and~(\ref{Review25}), one obtains the general relation
\begin{equation}
c(t)=H(t)R(t)\label{Review12}
\end{equation}
This can be regarded as a characteristic equation of the Universe
in this model.

The mean value of~(\ref{Review27}) is $t_0\approx 12$~Gyr, from the relation~(\ref{Review11}),
one gets
\begin{equation}
R_0\approx 15~\mathrm{Gly}\label{Review13}
\end{equation}
where the light-year unit is based on $c_0$. However, there is no
contradiction, for $c$ is time dependent and its value at previous
epochs was greater than now.

If one tries to construct a VSL-type theory, endowed with the
Robertson-Walker metric, in which the Universe is causally
connected, then, if $R\propto a\propto t^\lambda$ and $c\propto
t^\nu$, one must have
\begin{equation}
\lambda=\nu+1\label{Review14}
\end{equation}
Assuming that $\rho c^2\propto a^{-3}$ and the Universe, at least
for large $t$, is not curvature-dominated, from the first
Friedmann equation~(\ref{Semi3}), one can easily find that
$\lambda=\frac{4}{5}$ and $\nu=-\frac{1}{5}$. This is exactly what
has been derived in this model.

The total current mass of the Universe can be calculated from
relation~(\ref{Brief3}) as
\begin{equation}
M_0=\frac{R_0c_0^2}{G}\sim 10^{53} \mathrm{kg}\label{Review15}
\end{equation}
where the traditional value of the gravitational constant has been
used. Consequently, one obtains the energy content of the Universe as
\begin{equation}
{\cal E}_0=M_0c_0^2\sim 10^{70} \mathrm{J}\label{Review16}
\end{equation}
which is a constant all over the life of the Universe by our first
assumption.

To evaluate the current density of the Universe, one can obviously
suppose that the volume of it, when its radius is
$R_0$, is
\begin{equation}
V_0=\frac{4}{3}\pi \beta R_0^3\label{Review17}
\end{equation}
where $\beta$ is introduced to present the geometry of the
Universe. In other words, $0<\beta<1$, $\beta=1$, and $\beta>1$
correspond to $k=+1$, $k=0$, and $k=-1$, respectively. Now, by
relations~(\ref{Review15}) and~(\ref{Review17}), one has
\begin{equation}
\rho_0=\frac{3c_0^2}{4\pi \beta GR_0^2}\label{Review18}
\end{equation}

On the other hand, one can rewrite the first Friedmann
equation~(\ref{Semi3}) as
\begin{equation}
\Omega_0-1=\frac{kc_0^2}{a_0^2H_0^2}\label{Review19}
\end{equation}
where $\Omega_0\equiv\frac{\rho_0}{{\rho_c}_0}$ and
$\rho_c=\frac{3H^2}{8\pi G}$. Substituting
relation~(\ref{Review18}) in the above relation and using
relation~(\ref{Review24}), it yields
\begin{equation}
\frac{2}{\beta}-1=k\left(\frac{R_0}{a_0}\right)^2\label{Review20}
\end{equation}
Hence, the only possible values of $\beta$, in agreement with the
sign of $k$, are
\begin{equation}
\left\{
\begin{array}{lllll}
0<\beta<1 & \textrm{ where} & k=+1 & \textrm{ \&} & \Omega_0>2 \\
\beta>2 & \textrm{ where} & k=-1 & \textrm{ \&} & \Omega_0<1\quad
.
\end{array}\right.
\label{Review21}
\end{equation}
Note that, no exact flat geometry is permitted in this model. In
addition, considering the current cosmological data, the second
possibility, i.e., the negative curvature, seems more probable.
Introducing $\Omega_k\equiv-k\left(\frac{R_0}{a_0}\right)^2$, one
can also write~(\ref{Review20}) as
\begin{equation}
\Omega_0+\Omega_k=1\label{Review22}
\end{equation}
where, both $\Omega_0$ and $\Omega_k$ are positive constants
according to the above discussion.

A more complete model, in which some other fundamental physical
constants are supposed to vary, will be introduced
later~\footnote{H. Shojaie and M. Farhoudi. A
varying-constant cosmology. Manuscript in preparation}

\end{document}